\documentclass[12pt]{article} 
\usepackage[english,russian]{babel}
\usepackage[koi8-r]{inputenc} 
\usepackage{cyr}
\usepackage{epsf}
\usepackage[dvips]{graphicx}
\usepackage{pazh}
\usepackage{float}
\usepackage{amssymb}
\tightenlines

\def\etal{{et~al.}}
\addto\captionsrussian{} 
\sloppypar

\begin{document}
\begin{flushleft}
{\small\it Astronomy Letters, 2018, Vol. 44, No. 10, pp. 579-587. }
\\
\hrulefill
\end{flushleft}

\baselineskip 21pt

\title{\bf Absorption of Photons from Distant Gamma-Ray Sources
      }

\author{\bf \hspace{-1.3cm}
         A. N. Popov\affilmark{}, 
         D. P. Barsukov\affilmark{*},
         A. V. Ivanchik\affilmark{}  
       }

\affil{
{\it Ioffe Institute, Saint-Petersburg, Russia }}

\vspace{2mm}

\sloppypar 
\vspace{2mm}
\noindent
Being the largest gravitationally bound structures in the Universe,
galaxy clusters are huge reservoirs of photons generated 
by the bremsstrahlung of a hot cluster gas. 
We consider the absorption of high-energy photons 
from distant cosmological gamma-ray sources 
by the bremsstrahlung of galaxy clusters. 
The magnitude of this effect is the third in order of smallness
after the effects of absorption by 
the cosmic microwave background and 
absorption by the extragalactic background light. 
Our calculations of the effect of absorption 
by the bremsstrahlung of galaxy clusters have shown
that this effect manifests itself
in the energy range 1--100 GeV 
and can be $\tau \sim 10^{-5}$. 



\vfill
\noindent\rule{8cm}{1pt}\\
{$^*$ E-mail: bars.astro@mail.ioffe.ru}

\clearpage

\section*{INTRODUCTION}
\noindent
The emission from active galactic nuclei (quasars, blazars, etc.) 
and gamma-ray bursts is of special
interest in astrophysics, because the spectra of these
sources contain information about the physical conditions and 
the composition of matter in the Universe
at various stages of its evolution. 
For example, the
spectra of distant blazars in the gamma-ray range
($E > 100$ GeV) are currently studied with highly
sensitive Cherenkov telescopes, such as H.E.S.S and VERITAS 
(Abramowski et al. 2013; Lenain 2014; Pfrommer 2013). 
In this case, it is very important
to take into account the factors leading to spectral distortions 
as high-energy photons propagate through intergalactic space.
The absorption of
gamma-ray photons with the production of electron
positron pairs when gamma-ray photons collide with
background photons plays the most important role
here (Ruffini et al. 2016). 
For gamma-ray photons with energies 
$E \sim 100 \mbox{ GeV} - 100 \mbox{ TeV}$  
this is mainly
the interaction with extragalactic background light (EBL) photons 
(Franceschini et al. 2008; Sinha et al. 2014; Dwek and Krennrich 2013). 
In the energy range 
E  100 TeV10 7 TeV 
$E \sim 100 \mbox{ TeV} - 10^{7} \mbox{ TeV}$  
the interaction
with cosmic microwave background (CMB) photons dominates 
(De Angelis et al. 2013; Gould and G.P. Schreder 1967b). 
At energies $E > 10^{7} \mbox{ TeV}$
the interaction with cosmic radio background (CRB) photons 
becomes important 
(De Angelis et al. 2013; Gould and Schreder 1967b; Reesman and Walker 2013).

Galaxy clusters are the largest gravitationally
bound objects in the Universe. An appreciable
fraction of the cluster baryonic matter is 
a tenuous ($n \sim 10^{-3}-10^{-2} \mbox{ cm}^{-3}$) 
hot ($T \sim 1-10 \mbox{ keV}$) intracluster gas 
(see, e.g., Vikhlinin et al. 2006; Patej and Loeb 2015). 
The scattering of photons by particles of
this gas leads to distortions in the spectra of distant
sources (Sunyaev and Zeldovich 1972). In addition,
due to the bremsstrahlung of electrons in this gas,
clusters also turn into huge ($R \sim 1-10 \mbox{ Mpc}$) reservoirs
of photons (Lea et al. 1973). Therefore, apart from the
extragalactic background components listed above,
distortions in the spectra of distant cosmological
gamma-ray sources can also be caused by the interaction of 
gamma-ray photons with bremsstrahlung
photons of the hot gas in galaxy clusters.

In this paper the effect of photon absorption
with the production of an electronposition pair on
bremsstrahlung is calculated and compared with
the effect of absorption by the cosmic microwave
background.

\begin{figure}[t]
\includegraphics[width=0.90\hsize]{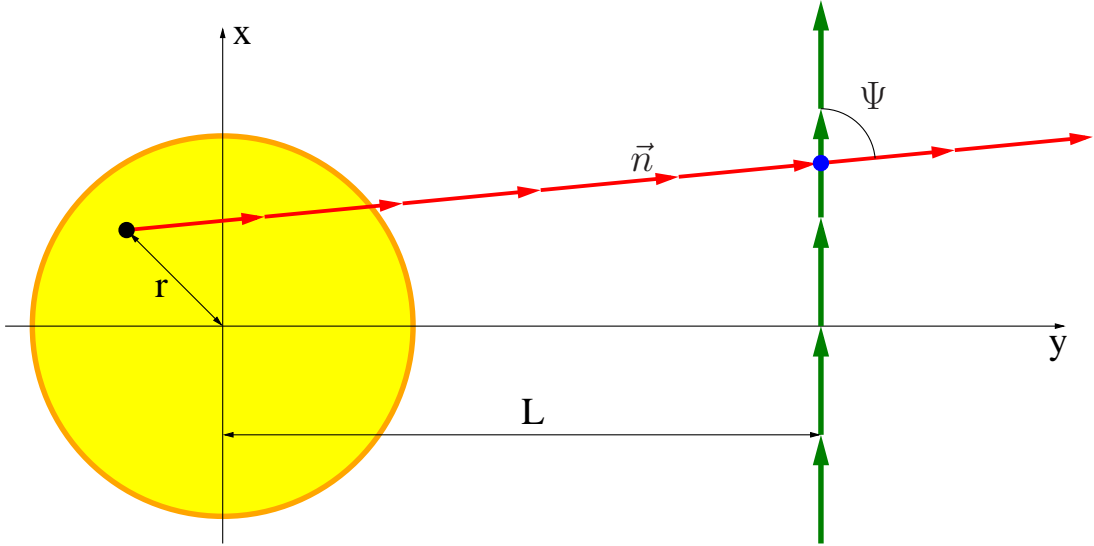}
\ \\ \ \\ \ \\
\caption{\rm 
 Scheme of the geometry under consideration. 
 The gas in the cluster is indicated by the gray circle; 
 the gamma-ray photon trajectory is indicated by the vertical straight line.
}
\label{fig_geom}
\end{figure}

\begin{figure}[t]
\includegraphics[width=0.80\hsize]{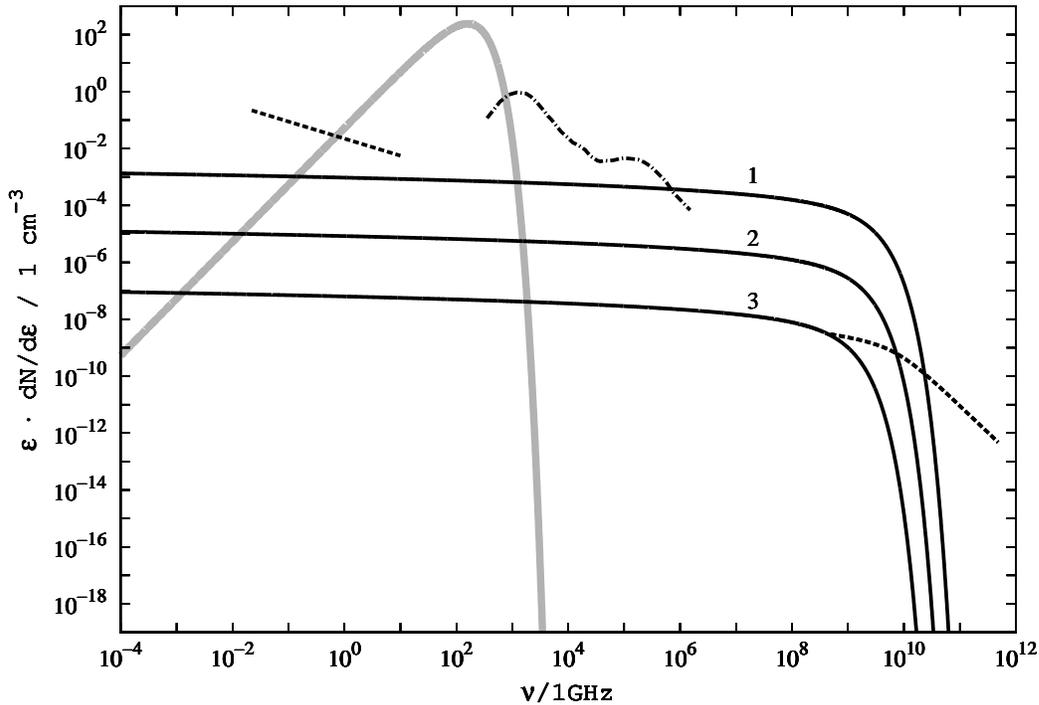}
\ \\ \ \\ \ \\
\caption{\rm 
The bremsstrahlung spectra used are indicated by the black solid curves. 
${dN}/{d\epsilon}$ is the number of photons per unit
volume per unit energy interval. 
Curve 1 corresponds to 
$n_{e}=10^{-2} \mbox{ \  cm}^{-3}$, $T_{e}=8 \mbox{ keV}$, $R = 7 \mbox{ Mpc}$;
curve 2 corresponds to 
$n_{e}=10^{-3} \mbox{ \  cm}^{-3}$, $T_{e}=5 \mbox{ keV}$, $R = 5 \mbox {Mpc}$;
and curve 3 corresponds to 
$n_{e}=10^{-4} \mbox{ \  cm}^{-3}$, $T_{e}=3 \mbox{ keV}$, $R = 3 \mbox{ Mpc}$.
The spectra used (for $z=0$) are also shown: 
the gray solid, back dashdotted, and
gray dotted lines indicate the CMB, EBL, and CRB spectra, respectively. 
}
\label{fig_sp}
\end{figure}

\begin{figure}[t]
\includegraphics[width=0.80\hsize]{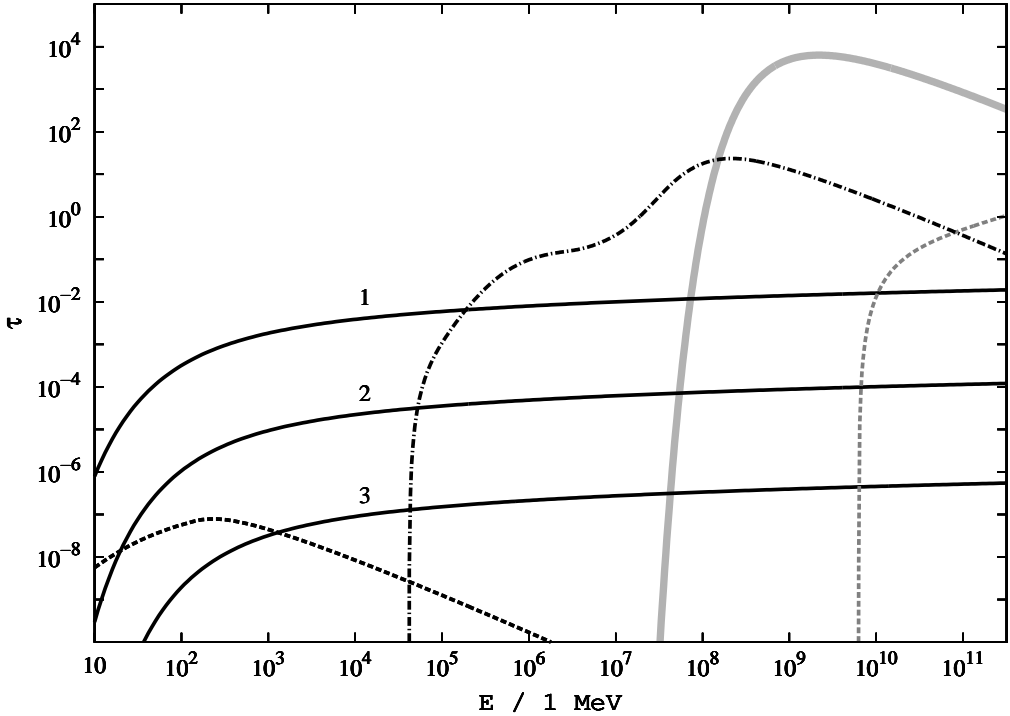}
\ \\ \ \\ \ \\
\caption{\rm 
Optical depth $\tau$ for a gamma-ray photon with energy $E$ 
corresponding to the spectra shown in Fig. \ref{fig_sp},
$z_{s} = 0.01$ and $z_{c} \ll 1$.                           
}
\label{fig_tau}
\end{figure}
\begin{figure}[t]
\includegraphics[width=0.80\hsize]{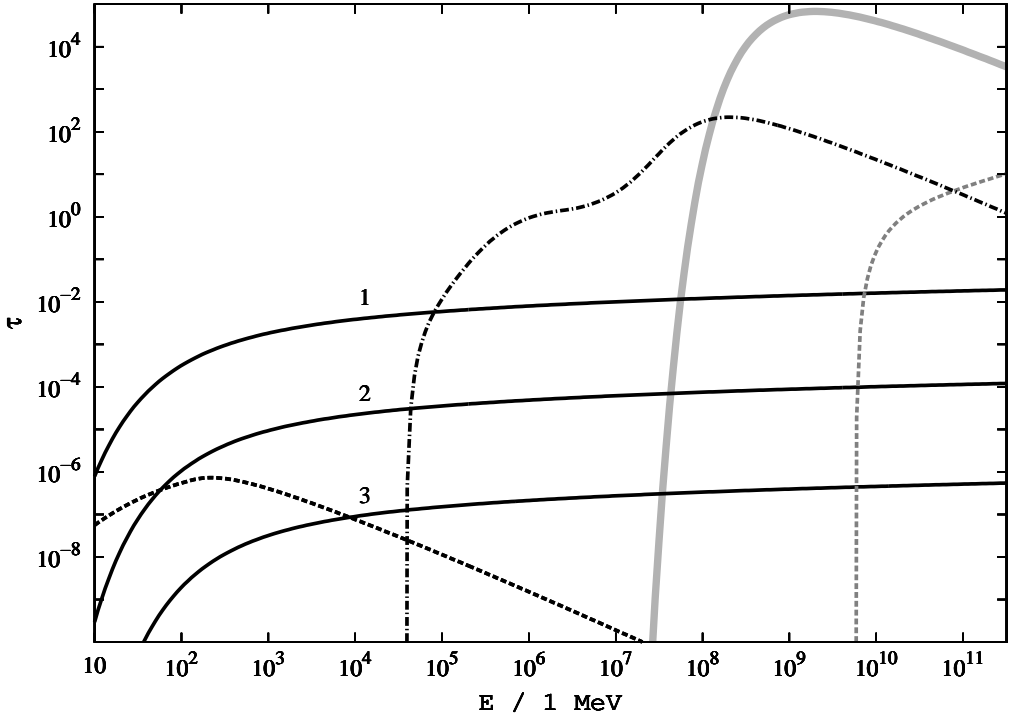}
\ \\ \ \\ \ \\
\caption{\rm 
Same as \ref{fig_tau} for $z_{s} = 0.1$ and $z_{c} \ll 1$.
}
\label{fig_tau_z1m1}
\end{figure}

\begin{figure}[t]
\includegraphics[width=0.80\hsize]{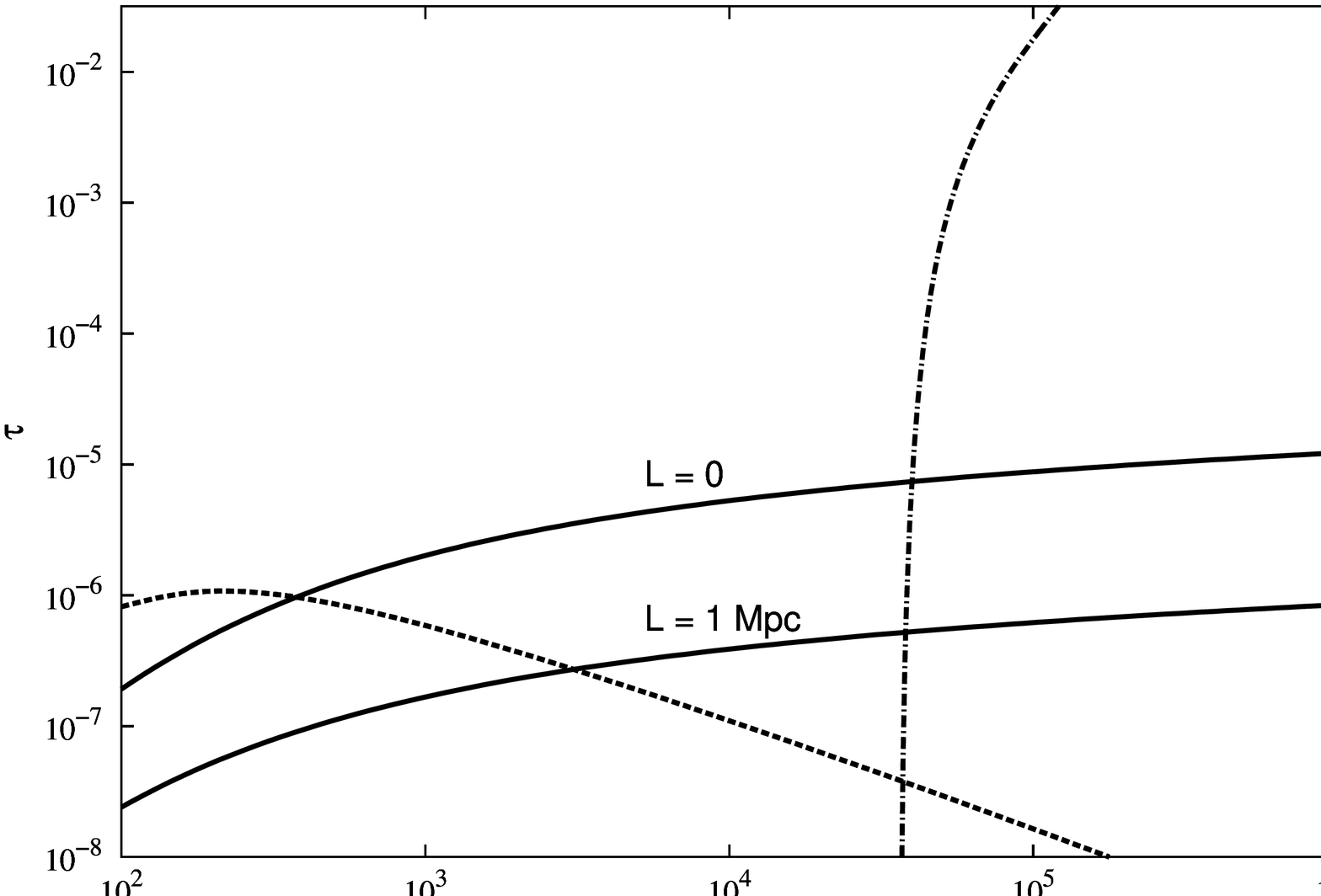}
\ \\ \ \\ \ \\ 
\caption{\rm
Optical depth $\tau$ versus gamma-ray photon energy $E$ 
for the cluster Abell 2204.    
}
\label{fig_basu_G}
\end{figure}

\section*{INTERACTION OF GAMMA-RAY EMISSION WITH THE EXTRAGALACTIC BACKGROUND}
\noindent  
Consider the interaction between two photons
with the production of an electron-positron pair:
\begin{equation}
\gamma_{E} + \gamma_{\epsilon} \longrightarrow e^{+} + e^{-}, 
\end{equation} 
where $\gamma_{E}$ is the incident gamma-ray photon with energy $E$ 
and $\gamma_{\epsilon}$ is the background photon 
with energy $\epsilon$. Its cross section is given by the
following expression (Gould and Schreder 1967a):
\begin{equation}
\sigma = \pi r^2_e \, \frac{1-\beta^2}{2} \,
    \left[   (3-\beta^4) \ln\left( \frac{1+\beta}{1-\beta} \right)
           - 2 \beta (2-\beta^{2})
    \right]
     H(s-1),
\label{eqn_1}
\end{equation} 
where $ r_e = {e^2}/{(mc^2)}$ is the classical electron radius,
$m$ is the electron rest mass,
$\beta ={v}/{c} $ is the center-of-mass velocity of 
the electron and positron divided by
the speed of light, 
and $H(x)$ is the Heaviside step function, 
$H(x) = 1$ at $x > 0$ and $H(x) = 0$ at $x \leq 0$.  
The relations between the center-of-mass velocity $\beta$  
of the produced particles, the photon energies $E$ and $\epsilon$, 
and the angle $\theta$ between the photon propagation
directions  are defined by the following equations
(Gould and Schreder 1967a):
\begin{equation}
s = \frac{\epsilon E}{2 m^2 c^4} (1-\cos {\theta})
\mbox{\hspace{1cm},  \hspace{1cm}}
\beta = \sqrt{1-\frac{1}{s}}.
\label{eqn_2_3}
\end{equation}
 
Consider a high-energy photon (with energy $E$) prop-
agating in a medium with low-energy photons. 
The gamma-ray photon absorption probability per unit
length at point $\vec{x}$ is (Gould and Schreder 1967a)
\begin{equation}
\frac{d\tau}{dx}(\vec{x})
  = \int_{-\infty}^{+\infty} 
            \sigma(E,\epsilon,\Psi)  
            (1 - \cos(\Psi)) 
            f(\vec{x},\vec{p}),
    d^3 p
\label{eqn_7}
\end{equation} 
where $f(\vec{x},\vec{p})$ is the distribution function of 
the low-energy photons, $\vec{p}$ is their momentum, and
$\epsilon = p c$. 
Let us integrate this expression along the line of sight
from $z=0$ (the cosmological redshift corresponding
to the present epoch and the time of signal observation) 
to $z = z_{s}$ (the cosmological redshift of the gamma-ray source). 
The optical depth  will then be
\begin{equation}
\tau=\int_0^{z_{s}} \,
     \int_{-\infty}^{+\infty} 
            \sigma(E,\epsilon,\Psi)  
            (1 - \cos(\Psi)) 
            f(\vec{x},\vec{p})  \frac{dx}{dz} \, 
     d^3 p \,  dz, 
\label{eqn_8}
\end{equation} 
where all quantities are taken on the photon trajectory and 
the factor ${dx}/{dz}$ describes the dependence of the
distance on redshift. 
In the standard $\Lambda CDM$-cosmological model 
it has the following form 
(at $z \ll z_{\rm eq} = 3365$, Ade et al. (2016), 
where the radiation
contribution $\Omega_{r}$ may be neglected):
\begin{equation}
\frac{dx}{dz} = \frac{c}{H_0} 
\frac{1}{(1+z)\sqrt{\Omega_{\Lambda}+\Omega_{M}(1+z)^3 }},
\label{eqn_6}
\end{equation} 
where $\Omega_{\Lambda}=0.692$ and $\Omega_{M}=0.308$
are the relative fractions of dark energy and dark matter, 
respectively (in units of the critical density),
$H_{0}= 67.8 \mbox{\ km}/\mbox{s Mpc}$
is the Hubble constant at the
present epoch (Ade et al. 2016).

\section*{INTERACTION OF GAMMA-RAY EMISSION
          WITH THE BREMSSTRAHLUNG OF GALAXY CLUSTERS
         }
\noindent
Let us estimate the hot-gas bremsstrahlung power
in a galaxy cluster. When a fast charged particle
collides with an ion, it undergoes deceleration and
emits bremsstrahlung photons. For a plasma that
has a Maxwellian velocity distribution of electrons
with temperature $T_{e}$ the emissivity of matter $\varepsilon_{\nu}$
can be written as (Lang 1980)
\begin{equation}
\varepsilon_{\nu}=\frac{8}{3}
    \left( \frac{2 \pi}{3} \right)^{1/2} \,
    \frac{Z^{2} \, e^6}{m^2c^3} \,
    \left( \frac{m}{kT_{e}} \right)^{1/2} \,
    n_{i} \, n_{e} \,
    g(\epsilon, T_{e}) \,
    \exp\left( -\frac{\epsilon}{kT_{e}} \right),
\label{eqn_10}
\end{equation} 
where $\nu$ is the bremsstrahlung photon frequency,
$\epsilon = h \nu$, $Z$ is the ion charge 
(for simplicity we will assume that $Z=1$),
$n_{e}$ and $n_{i}$ are the electron and ion densities, respectively,
$n_{i} = n_{e}$, 
$T_{e}$ is the electron temperature,
$g(\epsilon, T_{e})$ is the Gaunt factor 
(see, e.g., Zheleznyakov 1997):
\begin{equation}
g(\epsilon , T_{e}) = \frac{\sqrt{3}}{\pi} \,
  K_{0}(\epsilon / 2kT_{e}) \, 
  \exp\left( \frac{\epsilon}{ 2kT_{e}} \right),
\label{eqn_13}
\end{equation} 
where $K_{0}(z)$ is a Macdonald function.

For simplicity we will consider only small old relaxed clusters 
with a spherically symmetric gas distribution. 
We will assume that $z$ is virtually constant on the cluster scales. 
When considering the passage of a gamma-ray photon near the cluster,
we can then use a flat metric. 
Therefore, we will
assume that near the cluster the gamma-ray photon
propagates along a straight line
$\vec{x}(s) = L \vec{e}_{y} + s \vec{e}_{x}$
(Fig. \ref{fig_geom}).  
Consequently, supposing that the density of
gas bremsstrahlung photons in the cluster rapidly decreases
we can write
\begin{equation}
\tau\left( \frac{E}{1+z_{c}}\right) 
        = \int_{-\infty}^{+\infty} \,
          \int_{-\infty}^{+\infty} \, 
                 \sigma( E, \epsilon, \Psi) 
                 (1 - \cos\Psi ) 
                 f( \vec{x}(s) , \vec{p} )
                 \, d^{3}p \, ds,
\label{eqn_tau_cluster1}
\end{equation}  
where $z_{c}$ is the cluster redshift.

We will also assume that the cluster is optically thin for bremsstrahlung.
This is a good approximation at $\epsilon \gg \epsilon_{T}$, 
where the frequency  $\nu_{T} = \epsilon_{T} / h$ 
can be estimated as (Lang 1980)
\begin{equation}
\nu_{T} \approx 5 \mbox{ MHz} \, 
 \left( \frac{ T_{e} }{ 1\mbox{ keV} } \right)^{-0.675} 
 \left( \frac{ n_{e} }{ 1 \mbox{cm}^{-3} } \right) 
 \left( \frac{ \ell }{ 1 \mbox{ Mpc } } \right)^{1/2},
\label{eqn_18}
\end{equation} 
where $\ell$ is the extent of the plasma-filled region. 
Accordingly,
\begin{equation}
f(\vec{x},\vec{p})  =  \frac{2\pi}{\hbar} \, \frac{ c^{2} }{ \epsilon^{3} } \,
    \int_{0}^{+\infty}
          \varepsilon_{\nu}(\vec{x} - \lambda \, \vec{n}, \vec{p} )
    d\lambda,
\label{eqn_f_optic_thin}
\end{equation} 
where $\vec{n} = \vec{p} / p$.
We will take into account the fact
that the electron density $n_{e}$ and the gas temperature $T_{e}$  
depend only on the distance $r$ to the cluster center
and that the gas radiates isotropically. 
Then,
\begin{equation}
\varepsilon_{\nu}(\vec{x},\vec{p}) 
= \int_{0}^{+\infty}  \,
   \varepsilon_{\nu}(r,\epsilon) 
   \delta(|\vec{x}| - r)
  dr.
\label{eqn_intr}
\end{equation} 
Substituting 
Eqs. (\ref{eqn_f_optic_thin}) and (\ref{eqn_intr}) 
into (\ref{eqn_tau_cluster1}) gives
\begin{equation}
\tau\left( \frac{E}{1+z_{c}} \right) 
        = \int_{0}^{+\infty} \, \int_{0}^{+\infty} \, \int_{0}^{\pi} \,
              \sigma( E, \epsilon, \Psi) 
              (1 - \cos\Psi ) 
              \frac{2 \pi }{ \hbar c } 
              \frac{ \varepsilon_{\nu}(r,\epsilon) }{ \epsilon }
              I(r,L) \,     
          d\Psi \, d\epsilon \, dr,                   
\label{eqn_tau_cluster2} 
\end{equation} 
where the function $I(r,L)$ is
\begin{eqnarray}
I(r,L) & = & 4 \pi \, r \, {\mathrm{arcsin}}\left( \frac{r}{L} \right)
\mbox{ \  at \ } r < L,
\nonumber
\\
I(r,L) & = & 2 \pi^{2} \, r
\mbox{ \  at \ } r \geq L. 
\end{eqnarray}

\pagebreak
\begin{figure}[H]
\ \\
\includegraphics[width=0.80\hsize]{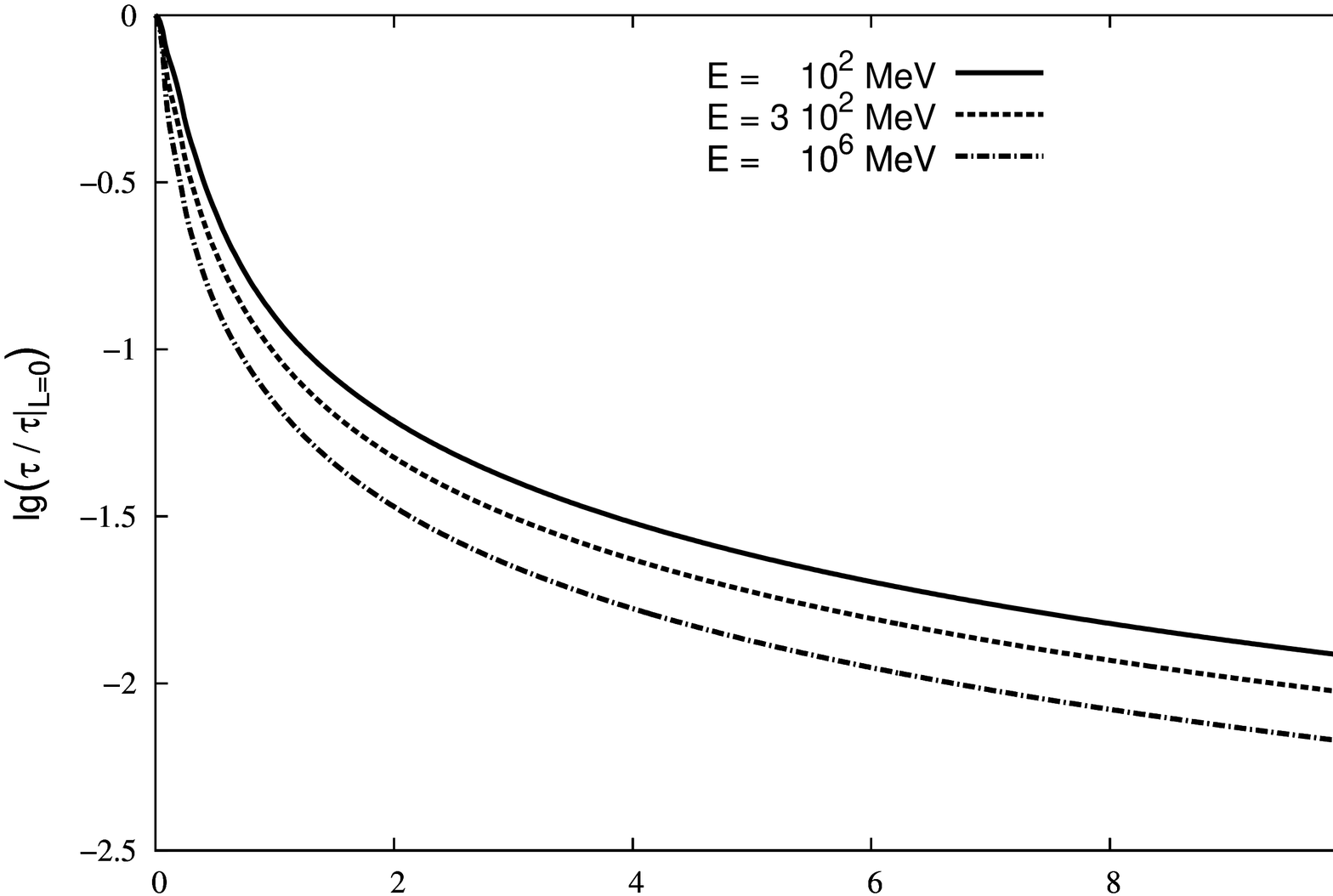}
\ \\ \ \\ \ \\ 
\includegraphics[width=0.80\hsize]{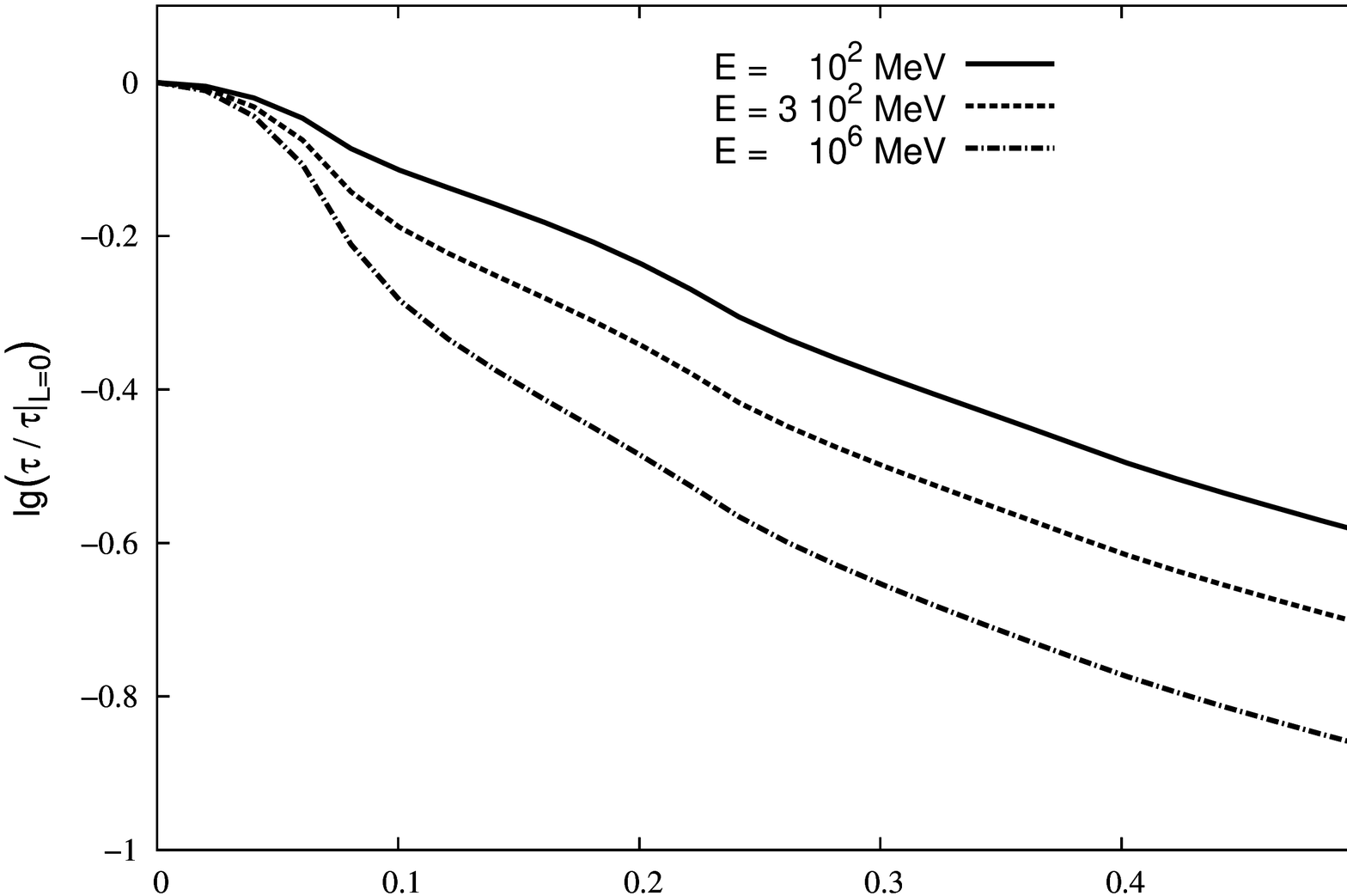} 
\ \\ \ \\ \ \\
\caption{\rm 
Normalized optical depth $\tau / \left. \tau \right|_{L=0}$
versus impact parameter $L$ 
for the cluster Abell 2204. 
The graphs differ only by the scale.   
}
\label{fig_basu_L}
\end{figure}

\pagebreak 
\begin{figure}[H]
\ \\
\includegraphics[width=0.80\hsize]{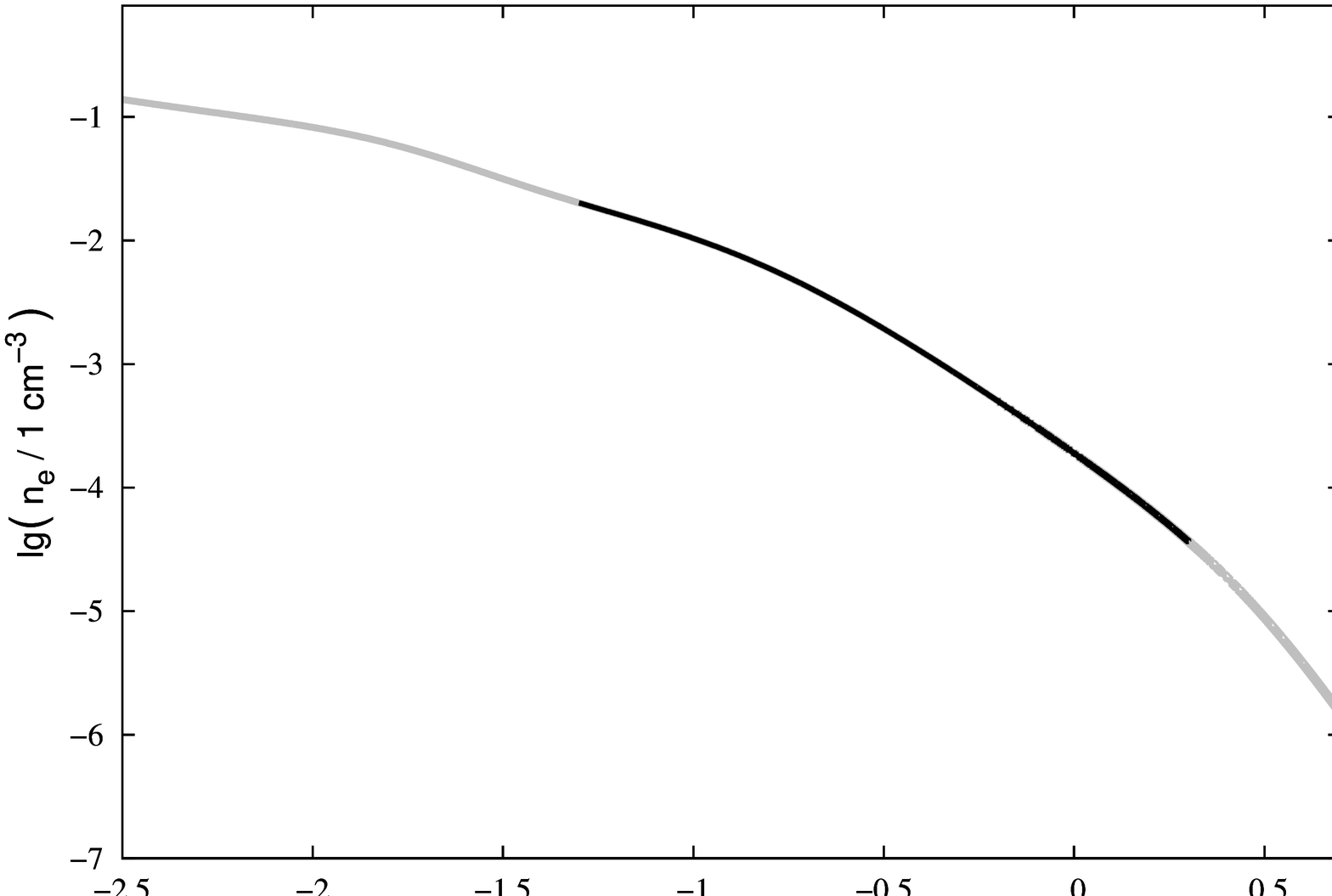}
\ \\ \ \\ \ \\ 
\includegraphics[width=0.80\hsize]{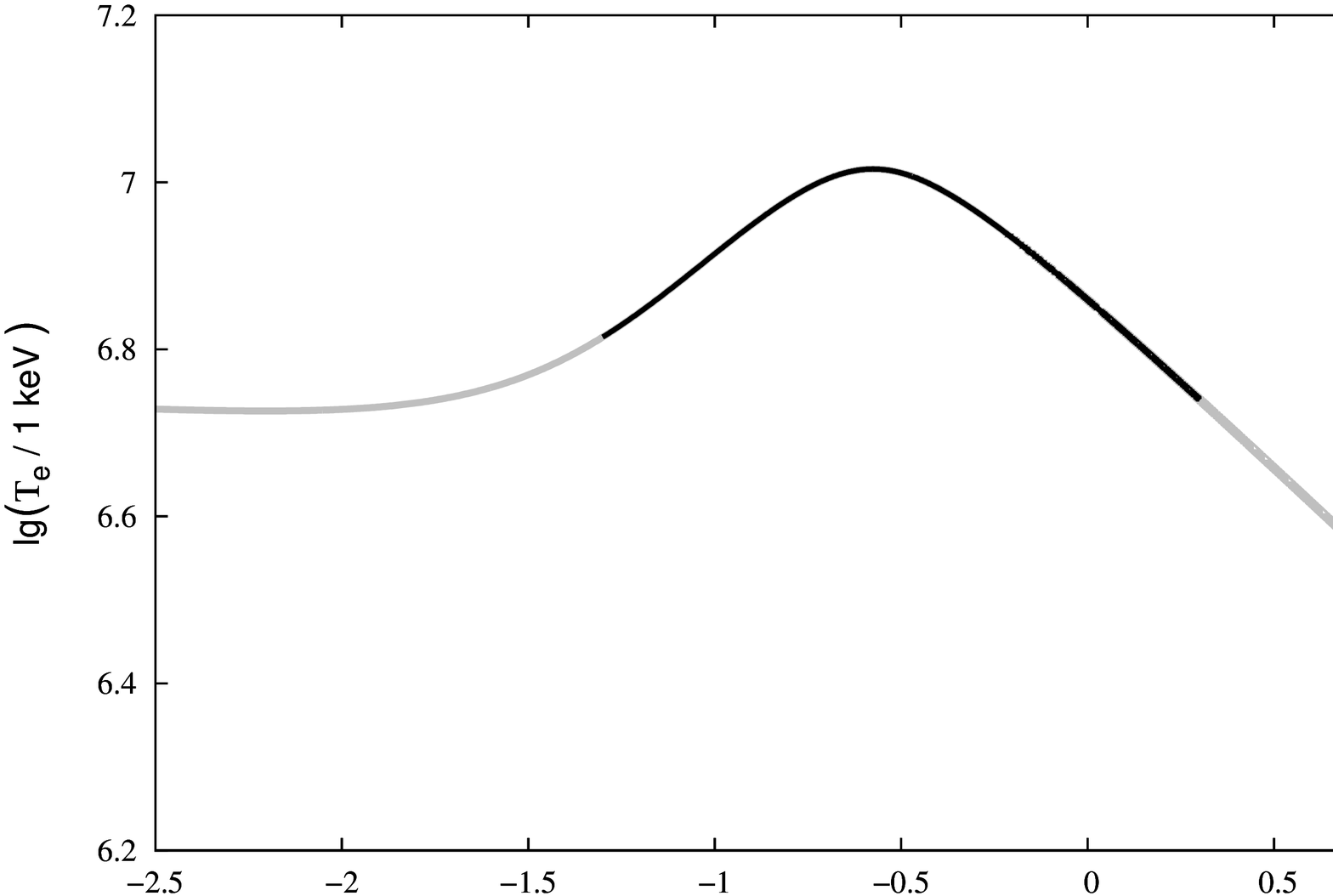} 
\ \\ \ \\ \ \\
\caption{\rm
Electron density $n_{e}(r)$ and gas temperature $T_{e}(r)$ 
versus distance $r$ to the cluster center for the cluster Abell 478: 
the black solid curves indicate the observed density 
and temperature profiles; 
the gray solid curves indicate their extrapolations.
The data were taken from Vikhlinin et al. (2006). 
}
\label{fig_a478_nT}
\end{figure}

\pagebreak
\begin{figure}[t]
\includegraphics[width=0.80\hsize]{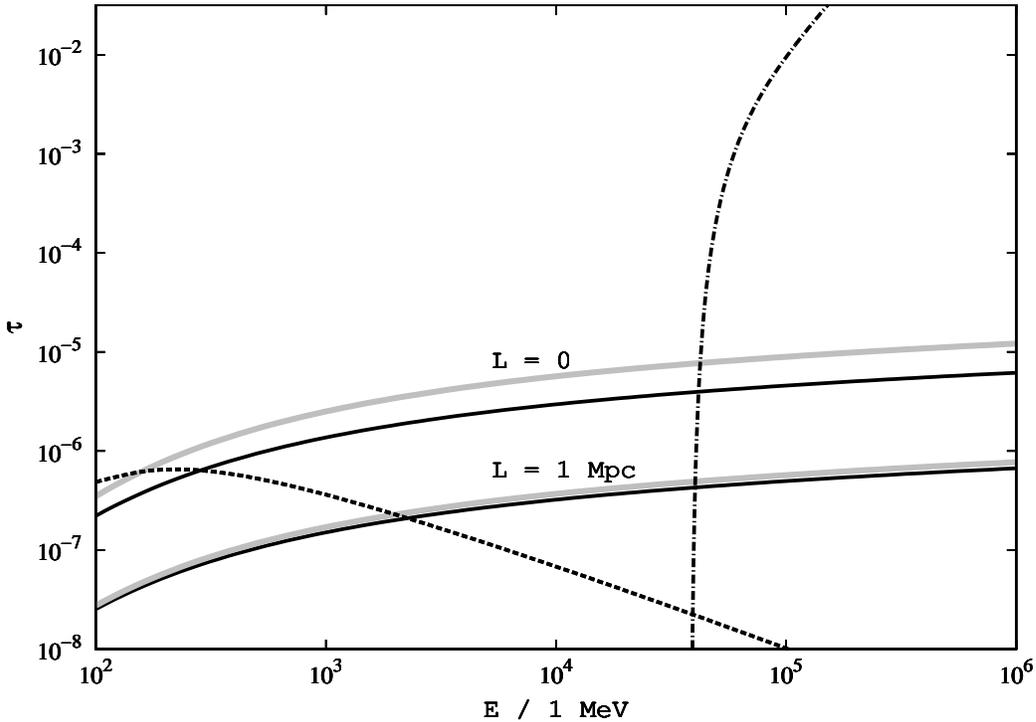}
\ \\ \ \\ \ \\ 
\caption{\rm
Same as Fig. \ref{fig_basu_G} for the cluster Abell 478: 
the black solid curve indicates the optical depth 
for the observed density and temperature profiles; 
the gray solid curve indicates the optical depth for
their extrapolations. 
}
\label{fig_a478_G}
\end{figure}

\section*{OPTICAL DEPTH}
\noindent
Consider first the simplest cluster model: 
a gas sphere of radius $R$ with a constant density $n_{e}$ and
temperature $T_{e}$, i.e., 
$n_{e}$ and $T_{e}$ are constant at $r<R$
and zero at $r > R$. 
Let us find the optical depths  $\tau$ for
three cases of galaxy cluster model parameters:

1) $n_{e}=10^{-2}$ cm$^{-3}$, $T_e=8$ keV, $R=7$ Mpc,

2) $n_{e}=10^{-3}$ cm$^{-3}$, $T_e=5$ keV, $R=5$ Mpc,

3) $n_{e}=10^{-4}$ cm$^{-3}$, $T_e=3$ keV, $R=3$ Mpc.
 
In Fig. \ref{fig_sp} the black solid curves indicate 
our estimates of the bremsstrahlung spectrum 
${dN}/{d\epsilon}$ 
at the cluster center $r=0$ for the cases listed above,
where ${dN}/{d\epsilon}$ is the number of photons 
with energy $\epsilon$ per unit volume per unit energy interval.
Figure \ref{fig_sp} also shows
the cosmic microwave background (CMB) spectrum
at $z=0$ (thick gray curve), 
the extragalactic background light (EBL) spectrum at $z=0$ 
taken from Franceschini et al. (2008) (black dashdotted curve),
the cosmic radio background (CRB) spectrum 
from Fixsen et al. (2011) (gray dashed curve), 
and the cosmic X-ray background (CXB) 
spectrum from Ajello et al. (2008) (black dashed curve). 
Figure \ref{fig_tau} presents
the optical depth $\tau(E)$ for absorption for these spectra 
provided that the gamma-ray source is at $z_{s} = 0.01$,
while the cluster is located, accordingly, at lower $z$. 
When calculating  $\tau$ for bremsstrahlung, 
we used Eq. (\ref{eqn_tau_cluster2}) and set $L = 0$.
In addition, since $z_{c} \leq 10^{-2} \ll 1$,
we neglected the photon redshift. 
Note that in this case $\tau$  does not depend on the specific
$z = z_{c}$ zbetween the source and the observer 
at which the cluster itself is located. 
The optical depth for the CMB, EBL, CRB, and CXB spectra 
was calculated from Eq. (\ref{eqn_8}) with $z_{s} = 0.01$.
The photon distribution function $f(\vec{x},\vec{p})$
was assumed to be isotropic. 
We also took into account the fact that the CMB temperature
in the standard cosmological model depends on $z$ as
$T = T_{0} \, (1+z)$, 
where $T_{0} = 2.725 \mbox{ K}$ is the CMB temperature 
at $z=0$ (Fixsen 2009). 
The EBL, CRB and CXB spectra were assumed to be independent of $z$.
Figure \ref{fig_tau_z1m1} shows the same quantities, 
but the source is at $z_{s} = 0.1$. 
Likewise, we neglected the redshift of the photons after 
their passage through the cluster,  $z_{c} \ll 1$, 
and assumed the EBL, RB, and CXB spectra to be independent of $z$.
Since we supposed that $T_{e}=const$, 
the normalized optical depth $\tau / \left. \tau \right|_{L=0}$ 
does not depend on the energy 
and depends only on the density profile. 
In the case of $n_{e}=const$
under consideration we have 
\begin{eqnarray}
\tau / \left. \tau \right|_{L=0} & = &
1 - \frac{1}{2} \frac{L^{2}}{R^{2}}
\mbox{\ at \ } L < R,
\nonumber
\\
\tau / \left. \tau \right|_{L=0} & = &
\frac{1}{\pi} \,
\left[
  \sqrt{ \frac{L^{2}}{R^{2}} -1 \, } \,
 - {\mathrm{arcsin}}\left( \frac{R}{L} \right) 
   \left( \frac{L^{2}}{R^{2}} - 2 \right)
\right] 
\mbox{\ at \ } L > R, 
\end{eqnarray} 
at $R \ll L$ this expression takes the form
\begin{equation}
\tau / \left. \tau \right|_{L=0} \approx 
\frac{4}{3} \, \frac{R}{L}. 
\end{equation}

\begin{table}[H]

\vspace{6mm}
\centering
{{\bf Table 1.} 
  Profiles of the electron density $n_{e}(r)$ 
  and gas temperature $T_{e}(r)$
  in the cluster Abell 2204. 
  The data were taken from Basu et al.(2010) 
}
\label{table_basu_nT} 

\vspace{5mm}
\begin{tabular}{llllllllll}
\hline\hline
$r$ & $1^{\prime}$  &
0--0.5 &  0.5--1.5 & 1.5--2.5 & 2.5--3.6 & 3.6--4.9 & 4.9--6.7 & 6.7--9.2 & 9.2--12.8
\\
$n_{e}(r)$ & $10^{-3} \mbox{cm}^{-3}$ &
24.4 & 5.30 & 1.79 & 0.86 & 0.47 & 0.27 & 0.11 & 0.06
\\
$T_{e}(r)$ & keV &
2.7 & 7.2 & 10.4 & 12.9 & 9.0 & 5.8 & 4.0 & 5.7
\\
\hline
\end{tabular}
\end{table}

Consider the gas in the cluster Abell 2204 
($z_{c}=0.1523$) (Basu et al. 2010). 
The dependences of the electron density $n_{e}(r)$ 
and gas temperature $T_{e}(r)$ on
the distance to the cluster center $r$ are given in 
Table 1 
(Basu et al. 2010). 
In our calculations we proceeded 
\pagebreak
\begin{figure}[H]
\includegraphics[width=0.80\hsize]{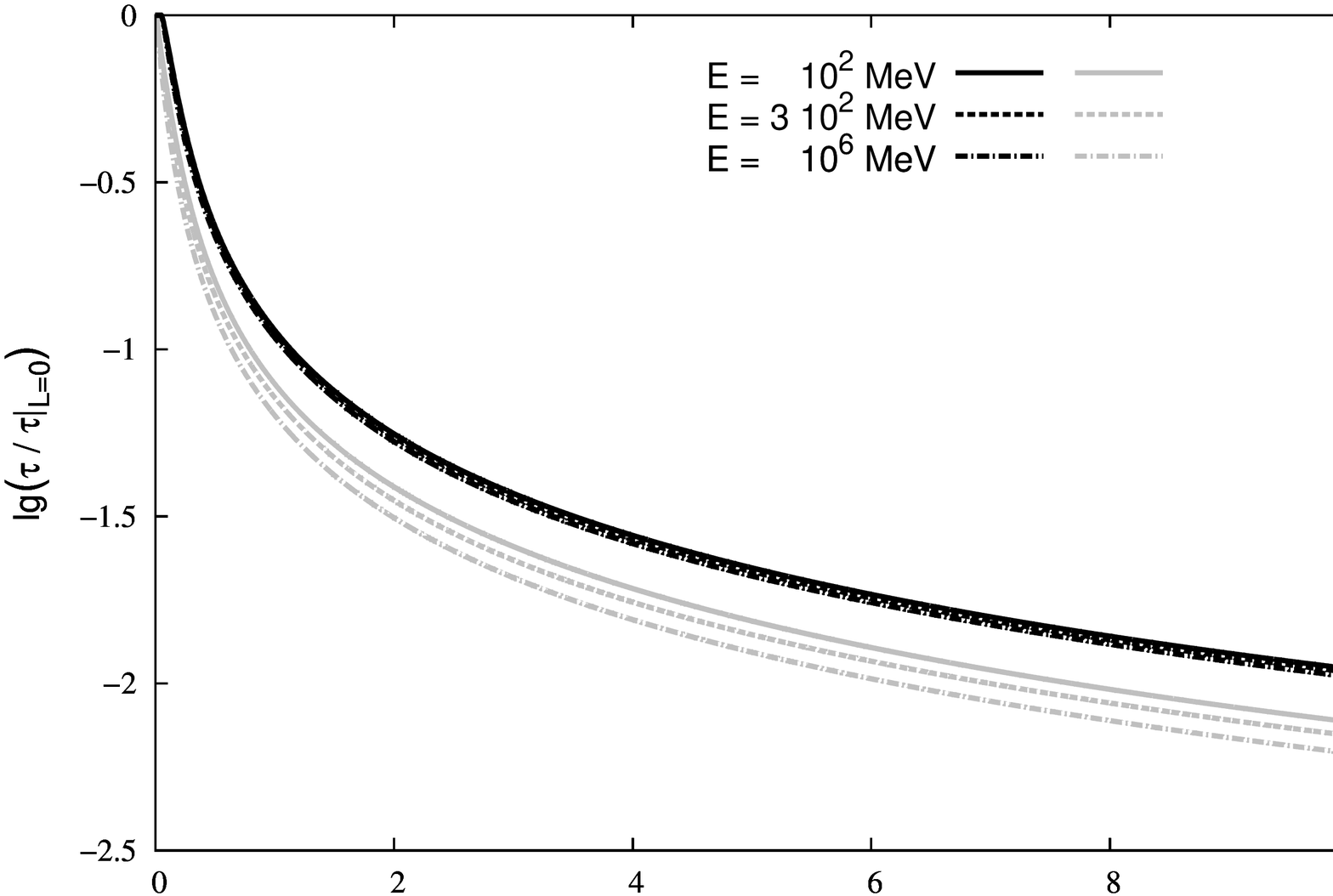}
\ \\ \ \\ \ \\ 
\includegraphics[width=0.80\hsize]{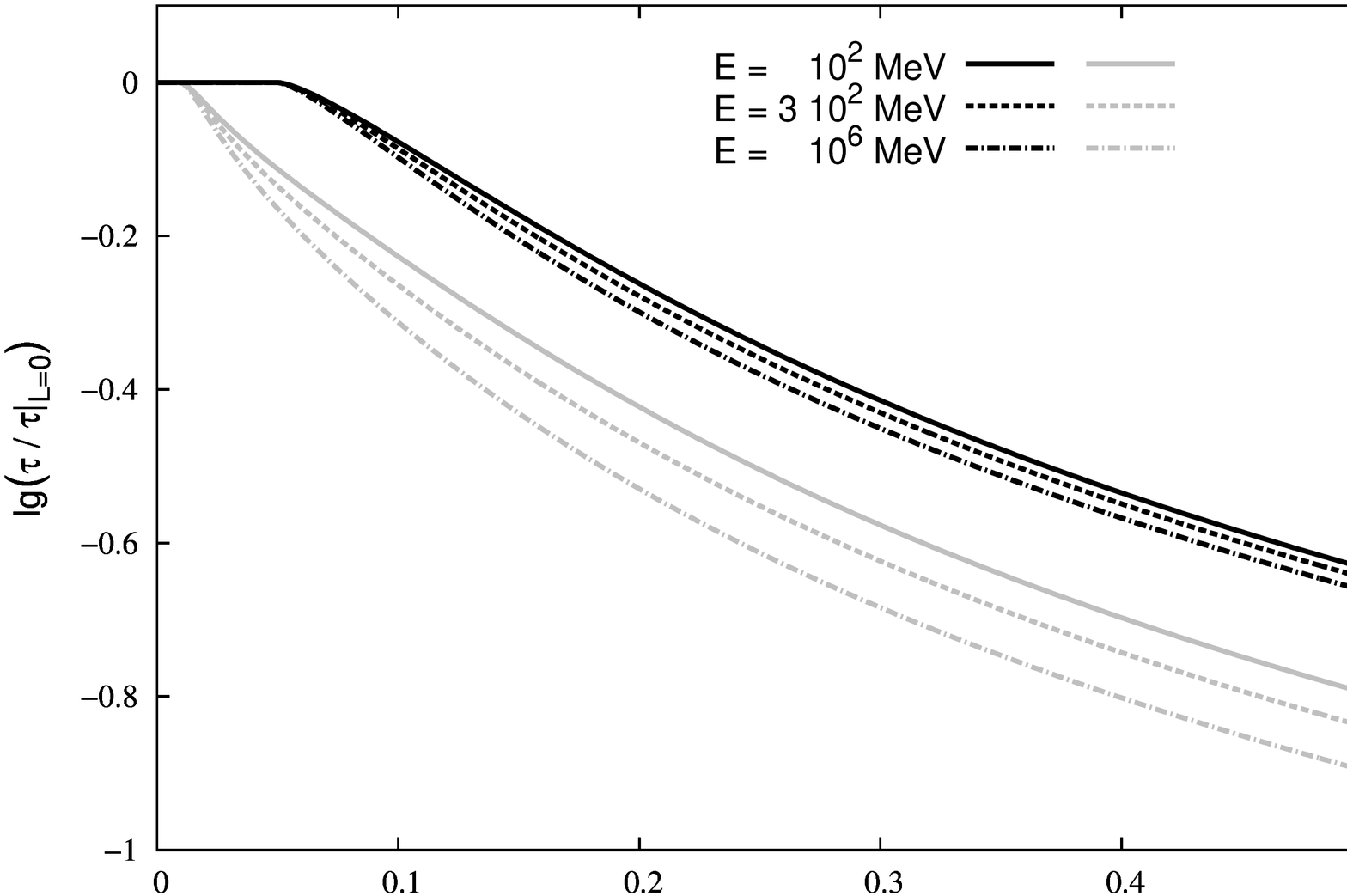} 
\ \\ \ \\ \ \\
\caption{\rm
Same as Fig. \ref{fig_basu_L} for the cluster Abell 478:  
the black solid curves indicate the optical depth 
for the observed density and temperature profiles; 
the gray solid curves indicate the optical depth for
their extrapolations. 
}
\label{fig_a478_L}
\end{figure}
\pagebreak
\noindent 
from the fact 
that the cluster radius $r_{200}=11.2^{\prime}$
corresponds to $1.76 \mbox{\ Mpc}$ (Basu et al. 2010).
In Fig. \ref{fig_basu_G} the black solid curves indicate 
the dependence of the optical depth $\tau(E)$ on photon energy $E$ 
calculated from Eq. (\ref{eqn_tau_cluster2}) 
for $L=0$ and $L=1 \mbox{\ Mpc}$. 
The optical depth corresponding to the interaction with EBL 
(black dashdotted curve) and CXB (black dashed curve)
photons assuming that the source is immediately 
behind the cluster $z_{s} = z_{c}$ is also shown. 
The interaction with EBL and CXB photons was calculated as 
for the case shown in Fig. \ref{fig_tau}. 
In Fig. \ref{fig_basu_L} the ratio $\tau / \left. \tau \right|_{L=0}$ 
is plotted against the impact parameter $L$ to the cluster center.

Consider now the cluster Abell 478 ($z_{c}=0.0881$)
(Vikhlinin et al. 2006). 
The electron density $n_{e}(r)$ and 
gas temperature $T_{e}(r)$ are plotted against 
the distance $r$ to the cluster center in Fig. \ref{fig_a478_nT}.  
The observed profile is indicated by the black solid line; 
its extrapolation is indicated by the gray solid line. 
Figure \ref{fig_a478_G} presents the dependence of 
the optical depth $\tau(E)$ on photon energy $E$
calculated from Eq. (\ref{eqn_tau_cluster2}). 
The black solid curves correspond to the optical depth for
the observed density and temperature profiles, while
the gray solid curves correspond to the optical depth
for their extrapolations. 
The curves for $L = 1 \mbox{ Mpc}$ virtually merge together. 
Figure \ref{fig_a478_G} also shows the optical depth 
in the interaction with background photons for $z_{s} = z_{c}$;
its calculations were carried out as for 
the case shown in Fig. \ref{fig_tau}. 
In Fig. \ref{fig_a478_L} the ratio
$\tau / \left. \tau \right|_{L=0}$ 
is plotted against the impact parameter $L$ to the cluster center. 
The black curves correspond to the optical depth 
for the observed density and temperature profiles, 
while the gray ones correspond to the optical depth 
for their extrapolations. 
The plateau at $L=0$ in Fig. \ref{fig_a478_L} stems from 
the fact that we disregarded 
the contribution of the cluster center
to the bremsstrahlung.

\section*{RESULTS}
\noindent
The spectra of distant sources must undergo
changes due to the influence of the absorption of photons 
with the production of electronpositron pairs.
As a result of the scattering by the cosmic microwave
background, the Universe is opaque for radiation
with an energy above 100 TeV. 
The second most intense type of background photons in the Universe
is presumably the extragalactic background light.
The cross section for the interaction of EBL photons
with gamma-ray photons reaches its maximum at
energies of the latter 100 GeV -- 1 TeV 
(De Angelis et al. 2013). 
At ultrahigh energies $E \sim 10^{6} \mbox{ TeV}$ 
the Universe is opaque due to the interaction with the
cosmic radio background (CRB).

We ascertained that the interaction with gas
bremsstrahlung photons in galaxy clusters makes
a minor contribution to the optical depth compared
to the absorption on CMB, EBL, and CRB photons
almost for all gamma-ray photon energies. 
However, we found that for energies 
$E \sim 1 \mbox{ GeV} - 100 \mbox{ GeV}$ 
this effect can dominate and is $\tau \lesssim 10^{-5}$.

\pagebreak   

\end{document}